\documentclass[prb,twocolumn,aps,showpacs,fixfloats]{revtex4}
\usepackage{graphicx}
\usepackage{bm}
\usepackage{amsmath,amssymb}
\usepackage{subfigure}
\usepackage{float}
\usepackage{latexsym}
\usepackage{epstopdf}
\usepackage{color}
\usepackage{enumerate}
\usepackage{pdfpages}

\newcommand{\s}{\scriptscriptstyle}

\begin{document}

\title {Signature of Hanle Precession in Trilayer MoS$_2$: Theory and Experiment}

\author{K. Tian$^{1}$, Z. Yue$^{2}$, D. Magginetti$^{1}$, M. E. Raikh$^{2}$, and A. Tiwari$^{1}$ }

 \affiliation{$^{1}$Department of Materials Science and Engineering, University of Utah, Salt Lake City, UT 84112, USA \\
$^{2}$Department of Physics and
Astronomy, University of Utah, Salt Lake City, Utah 84112, USA}

\begin{abstract}
Valley-spin coupling in transition-metal dichalcogenides (TMDs) can result in unusual spin transport behaviors under an external magnetic field. Nonlocal resistance measured from 2D materials such as TMDs via electrical Hanle experiments are predicted to exhibit nontrivial features, compared with results from conventional materials due to the presence of intervalley scattering as well as a strong internal spin-orbit field. Here, for the first time, we report the all-electrical injection and non-local detection of spin polarized carriers in trilayer MoS$_2$ films. We calculate the Hanle curves theoretically when the separation between spin injector and detector is much larger than spin diffusion length, $\lambda_s$. The experimentally observed curve matches the theoretically-predicted Hanle shape under the regime of slow intervalley scattering. The estimated spin life-time was found to be around $110~ps$ at $30$ K.
\end{abstract}

\maketitle

\begin{figure}
\includegraphics[width=84mm] {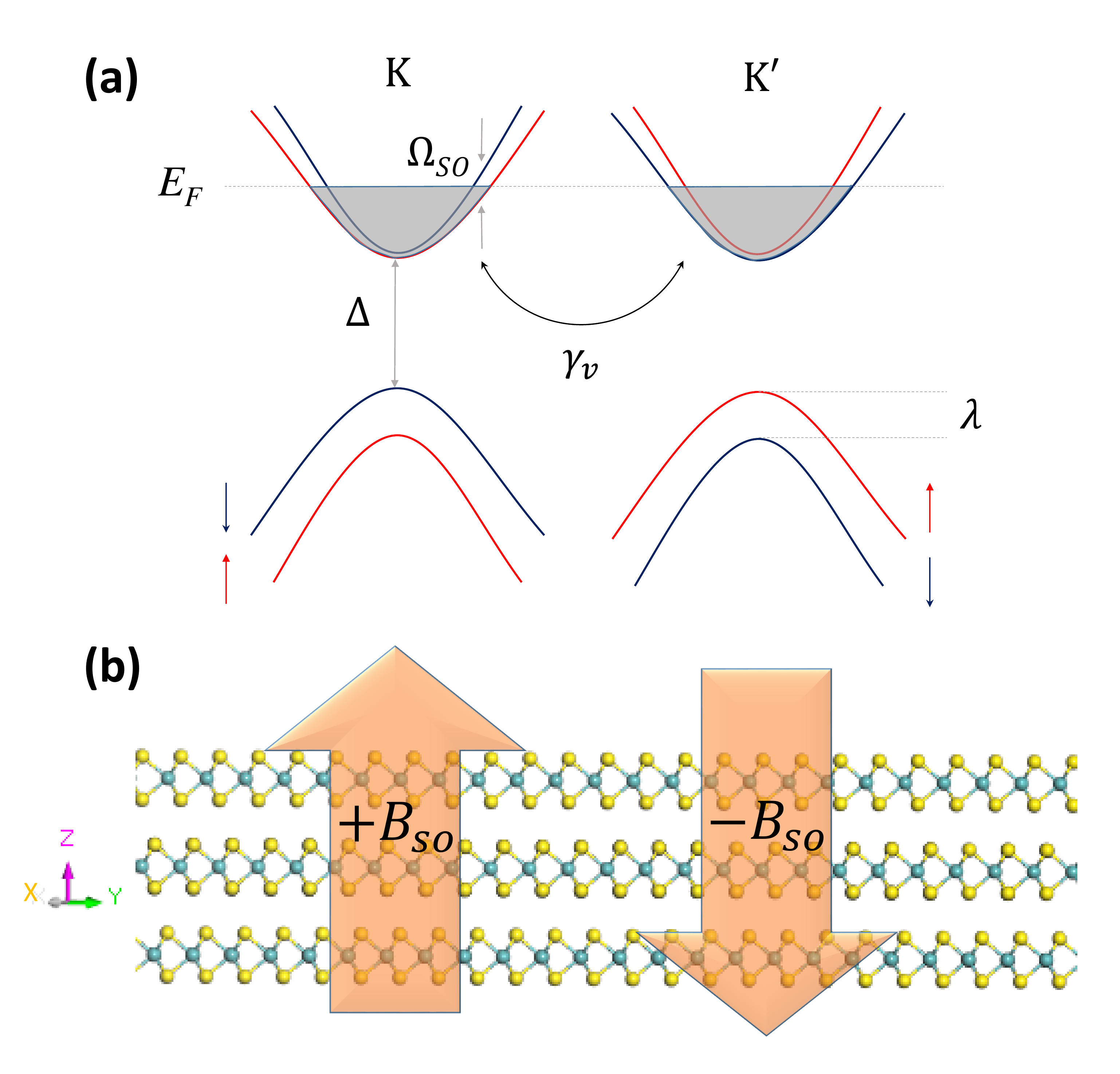}
\caption{(Color online) (a) Schematic of energy spectrum of trilayer MoS$_{2}$ at $K$ and $K'$ valleys (top). The spin splitting at the Fermi level is $\Omega_{\s SO}$=$\big(\frac{\lambda}{\Delta}\big)E_{\s F}$, where $\lambda$ is the spin splitting at valence band and $\Delta$ is the band gap. Intervalley scattering rate, $\gamma_v$, is determined by short range impurities.  Bottom scheme indicates the internal spin orbital field, $\pm B_{\s SO}$, induced by spin orbital coupling along z direction in the case of trilayer MoS$_2$.}
\label{fig1}
\end{figure}

\section{Introduction}
The discovery of two-dimensional (2D) transition metal dichalcogenides (TMDs) has attracted considerable attention recently due to their exotic properties, which are very important for their applications in electronic, optoelectronic, and spintronic devices [\onlinecite{1,2,3,4}]. Some members of the TMD family possess spin as well as valley degrees of freedom which make them attractive for next-generation quantum computing applications too. One of the most studied examples of above materials is MoS$_2$, which, contrary to graphene, possesses a band gap of $\Delta=1.6$ eV in its monolayer form [\onlinecite{5}]. The valleys in 2D MoS$_2$ are located at two energetically equivalent symmetry points, $K$ and $K'$, in the hexagonal Brillouin zone. Unlike graphene, the electron states in $K$ and $K'$ valleys are not equivalent. Because of large spin orbit coupling (SOC) which originates from the d orbitals of heavy Mo atoms [\onlinecite{8++,9,10,11}] and the inversion asymmetry induced Dresselhaus coupling [\onlinecite{7, 12, 13, 14}], the valence band edges of 2D MoS$_2$ undergo a large spin splitting ($\lambda \sim$150 meV). Furthermore, the Hamiltonian of 2D MoS$_2$ possesses time reversal symmetry, which leads the two valleys to exhibit opposite sign of spin-orbit field, see Fig.\ref{fig1} [\onlinecite{6, 7, 8}]. Because of the above unique characteristics, the spin and valley degrees of freedom in 2D MoS$_2$ can be controlled and manipulated independently, leading to its potential for applications in next-generation valleytronics-based devices.

Despite MoS$_2$'s great potential in future spintronic devices, there are still very few experimental studies on the spin transport and relaxation mechanisms operating in the material [\onlinecite{14+1,14+2,14+3}]. Among those, the most noteworthy study is by Yang \emph{et al.} [\onlinecite{15}] in which they used optical Hanle-Kerr experiment to measure the coupled spin-valley dynamics in 2D MoS$_2$ and observed a long spin lifetime of $3~ns$ at 5 K. Though the above observation is very important for obtaining fundamental understanding of the spin-valley dynamics, it is also important to explore spin transport characteristics of the material using all-electrical techniques.
Nonlocal Hanle techniques have been widely used to investigate spin transport in several semiconducting systems, see e.g. Refs. [\onlinecite{16, 19, Ian, 17, 18, 22,29}], however, there are no such reports for 2D MoS$_2$.

In nonlocal Hanle measurements, spin polarized carriers are injected into the semiconductor channel from a magnetized ferromagnetic electrode. Accumulated spin-polarization just below the injector electrode diffuses in the channel and creates spin imbalance below the  detector electrode.
This imbalance results in a voltage signal in the detector. However, when a transverse magnetic field is applied, the spin of the electrons starts precessing around the applied field and the voltage falls off.
The decay of voltage with magnetic field is referred to as Hanle curve. Its shape yields important information about the spin lifetime and spin diffusion length [\onlinecite{silsbee1, silsbee2}].

One reason which is  probably  responsible for the lack of reports on nonlocal Hanle studies on 2D MoS$_2$ is  that  large area films were not availabe until recently. Nonlocal Hanle experiments require four electrode contacts and at least two of those contacts must be long enough so that their shape anisotropy preserve their in-plane magnetization when a traverse magnetic field is applied. These experimental requirements are indeed quite challenging in the case of micron sized MoS$_2$ films normally produced by exfoliation based techniques. However, recently the growth of centimeter-scale high-quality 2D MoS$_2$ films by CVD and PVD techniques has been demonstrated by several groups [\onlinecite{3,26,CVD}]. Availability of these relatively large area films now can catalyze the spin transport studies on this exotic material system.

Other important factor, which possibly precluded the Hanle studies is the fact that 2D MoS$_2$ possesses very strong out-of-plane intrinsic magnetic fields. These built-in fields, which originate due to the lack of inversion symmetry in the monolayer MoS$_2$, can be of the order of several Tesla [\onlinecite{32}]. Since in electrical Hanle experiments the spin of the injected electrons is oriented in the plane of the channel, the intrinsic fields that are in the transverse direction can cause immediate precession and dephasing. In our previous theoretical study, we showed that the Hanle curve under normal field orientation exhibits  a two-peak structure with maxima located at the values of external field $B=\pm B_{\s SO}$,  where $B_{\s SO}$ is internal spin-orbit field. For monolayer MoS$_2$ the value of $B_{\s SO}$ is in the range of a few Tesla [\onlinecite{32}]. Thus, in experiments where the measurements are performed over a small out-of-plane external field interval, the peaks cannot be detected.


 While the strength of SOC in monolayer MoS$_2$ is expected to be the strongest, it is expected that SOC will also be present in other inversion asymmetric odd-layered MoS$_2$ films, see Ref.  [\onlinecite{13}]. The magnitude of SOC is supposed to decrease on increasing the number of
 layers. As a result, the two peaks will get progressively closer and might fall within the
 measurement range.

 With this motivation, we performed nonlocal Hanle measurements on trilayer MoS$_2$ films and obtained nontrivial results, which we report in this paper. To compare these results with theory, we have also extended our previous theoretical work Ref. [\onlinecite{25}] to incorporate the case of a finite distance between injector and detector.



The paper is organized as follows. In Section II, we  discuss the sample preparation, characterization and Hanle experiments. Section III describes a theoretical model used in our study. In Section IV, we present our experimental Hanle results and compare them with theory.
\begin{figure}
\includegraphics[width=80mm]{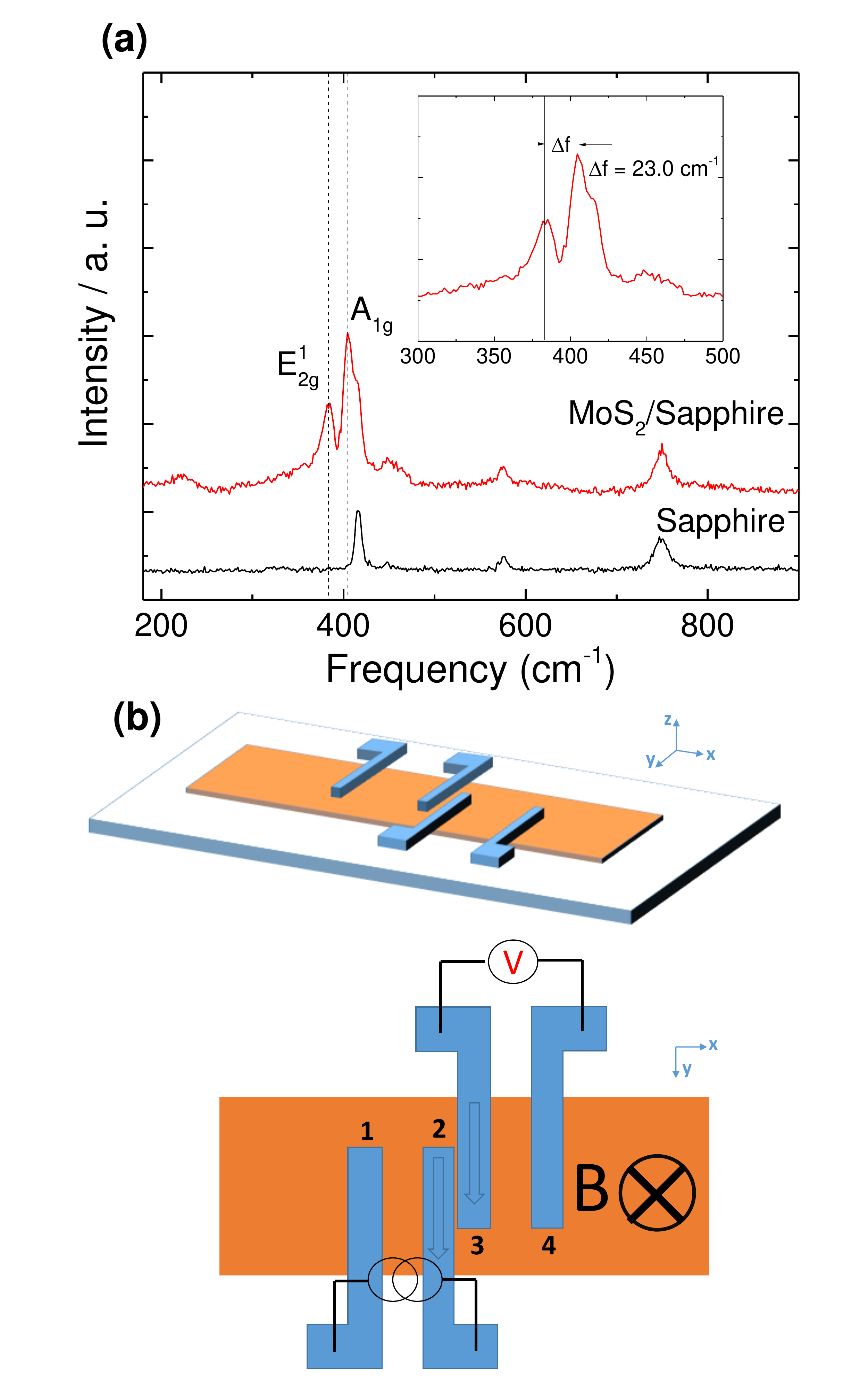}
\caption{(a) Raman spectra of trilayer MoS$_2$. The separation of vibration frequency $E_{2g}^{1}$ and $A_{1g}$ indicates three monolayers of MoS$_2$. (b) Schematic illustration of Hanle experiments performed through electrical methods on trilayer MoS$_2$, three dimensional view and top view. The measure of a nonlocal resistance is a voltage between right two contacts (3 and 4), while the current flows through the first two contacts (1 and 2), which are made of 15~nm NiFe. The distance of middle two contacts (2 and 3), $d$ (distance between the spin injector and detector) is 1~$\mu m$. We studied the Hanle curves under condition of $d > \lambda_{s}$. The arrows in the FM indicate the magnetization direction and also the spin polarization direction. Hanle experiments were performed under normally oriented external magnetic field.}
\label{fig2}
\end{figure}

\section{EXPERIMENTAL DETAILS}
Trilayer MoS$_2$ films were grown on sapphire substrates by pulsed laser deposition (PLD) following  the procedure described in previous report [\onlinecite{26}]. The number of monolayers deposited was controlled precisely by controlling the number of laser pulses. After deposition, Raman spectra were collected from the MoS$_2$ films as shown in Fig. \ref{fig2}(a). The two Raman vibrational modes, $E_{2g}^{1}$ and $A_{1g}$, confirm the presence of MoS$_2$.  In prior studies it has been demonstrated  that the separation between these modes can be used to determine the number of monolayers [\onlinecite{3}]. The observed peak separation of $23.0$ $cm^{-1}$ in the present study confirmed the formation of trilayer Mo$S_2$.

Electrical Hanle measurements were conducted using a four probe geometry as shown in Fig. \ref{fig2}(b). The pattern of ferromagnetic (FM) contacts was fabricated through photolithography, and electron beam evaporation was used to deposit $15$ nm thick NiFe contacts. After lift-off, the edge-to-edge separation of the middle two contacts was found to be 1 $\mu m$.

\begin{figure}
\includegraphics[width=78mm]{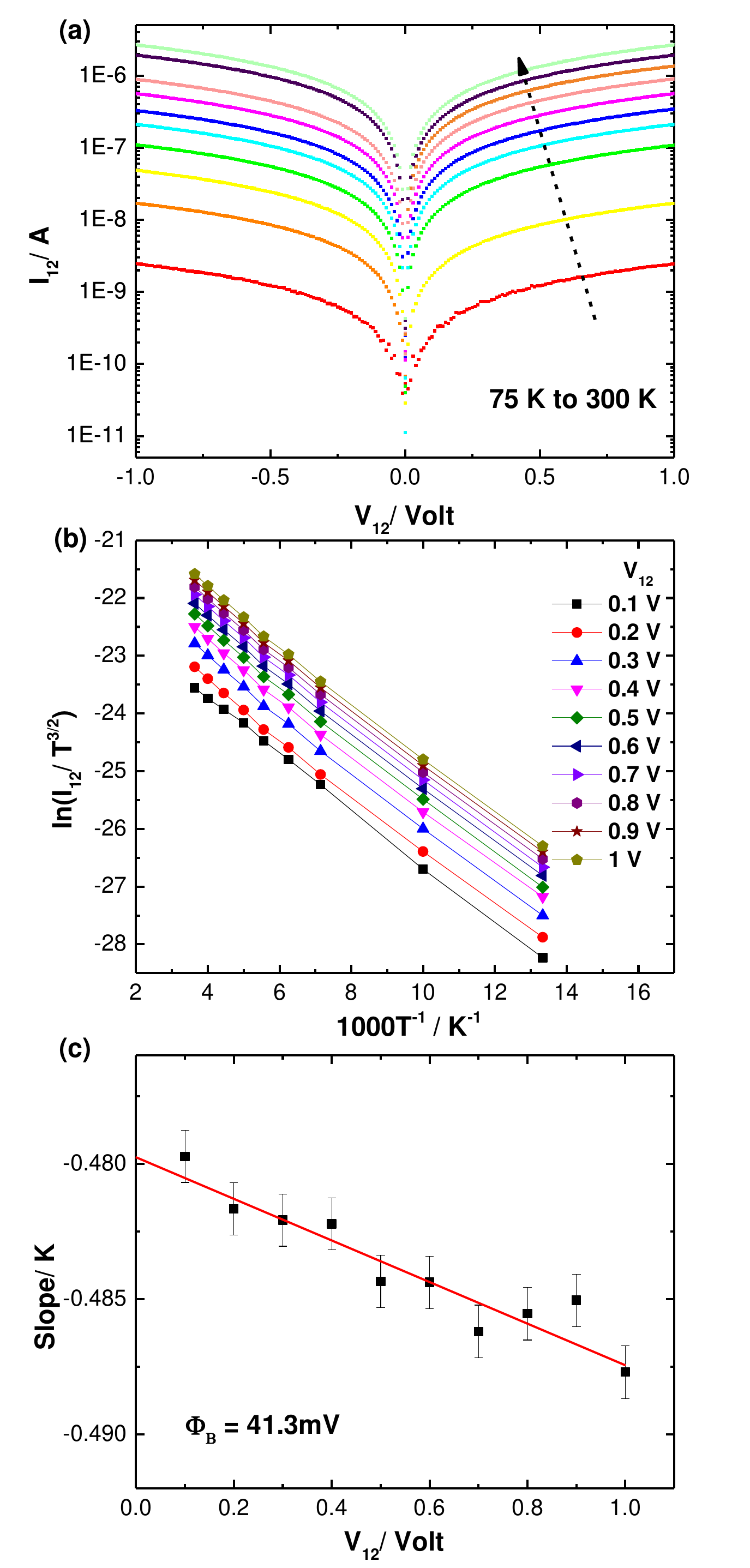}
\caption{(a) I-V curves of the trilayer MoS$_2$ sample with NiFe contacts from 75~K to 300~K. (b) Arrhenius plots, In (I$_{12}$/T$^{3/2}$) vs 1000~T$^{-1}$, at different V$_{12}$. (c) Extraction of Schottky barrier height,$\Phi_{\s B}$.}
\label{fig3}
\end{figure}

I-V measurements performed between contact 1 and 2 showed the presence of a Schottky barrier between NiFe and MoS$_2$. Presence of this barrier elinimated the need for depositing any additional tunnel barrier layer[\onlinecite{24, 27}]. To determine the Schottky barrier height ($\Phi_B$), temperature dependent I-V curves were recorded. Fig.\ref{fig3}(a) shows the I-V curves at different temperatures on a logarithmic scale. The barrier height was extracted using the thermionic emission function described in Appendix A. For this, first of all, $In(I_{12}/T^{3/2})$ vs. $1000/T$ was plotted for various $V_{12}$ values as shown in Fig. \ref{fig3}(b). In the next step, slopes of these plots were determined as a function of $V_{12}$ (see Fig. \ref{fig3}(c)). From the y-axis intercept of Slope vs. $V_{12}$ plot, value of Schottky barrier height was determined to be $41.3$ mV. For Hanle measurement, the current was passed between first two contacts, 1 and 2, while the non-local voltage, $V_{34}$, is measured through the other two contacts, 3 and 4.

\section{THEORY}

In the theory of the Hanle effect [\onlinecite{silsbee1, silsbee2, 28, 21, roundy}], the nonlocal resistance, $R(\omega_L^{z})$, is related to the spin density,  $S_x(t)$, as follows

\begin{equation}
\label{R}
R(\omega_L^{z}) = \frac{P^2 D \rho}{A}  \int\limits_0^\infty dt S_x(t) P_{d}(t),
\end{equation}
where $\omega_L^{z}$ is the Larmor frequency, $\omega_L^{z}$ = $\mu_{B} gB/\hbar$, $P$ is the spin injection/detection polarization, D is the diffusion coefficient related to the mobility ($\mu$) via the Einstein relation, $D=\mu k_{B} T/ e$, $\rho$ is the resistivity of channel material, $A$ is the cross-sectional area of the channel, and  $P_d(t)$ is the diffusion propagator defined as:


\begin{equation}
\label{diffusionpropogator}
P_d(t) = \frac{1}{(4\pi Dt)^{1/2}}\exp\Big(-\frac{d^2}{4Dt}\Big)
\end{equation}
In above equation, $d$ denotes the distance between the injector and detector electrodes.

The shape of the Hanle curve depends on the relation between $d$ and the spin diffusion length,
$\lambda_{s}$. In Ref. [\onlinecite{25}] we considered the case $d\ll \lambda_s$.
With regard to our present experimental study, the opposite limit $d \gg \lambda_s$ is relevant. Physically, in this limit, the Hanle curve is expected to exhibit numerous oscillations due to the fact that the spin of injected  electron can accomplish integer number of full precessions before it reaches the detector [\onlinecite{21}].

The unique characteristics of the spin dynamics in
 \newline TMDs originates from the fact that there exist two groups of
spins corresponding to two valleys $K$ and $K'$. Due to a finite intervalley scattering rate, $\gamma_v$, the time evolution of ${\bm S}^K(t)$ and ${\bm S}^{K'}(t)$ is described by the following system of {\em coupled} equations:
\begin{align}
\label{general}
\frac{d{\bm S}^K}{d t} &= (\Omega_{\s SO}+ \omega_L^{z}){\hat z} \times  {\bm S}^K
\hspace{-2mm} - \gamma_v \left( {\bm S}^K\hspace{-1.2mm} -\hspace{-0.7mm} {\bm S}^{K'} \right), \nonumber \\
\frac{d {\bm S}^{K'}}{d t} &= -(\Omega_{\s SO}- \omega_L^{ z}){\hat z} \times  {\bm S}^{K'}
\hspace{-2mm}+\gamma_v \left( {\bm S}^K\hspace{-1.2mm} - \hspace{-0.7mm}{\bm S}^{K'} \right),
\end{align}
where $\Omega_{\s SO}= \mu_{B} g B_{SO}/\hbar$. The above equations reflect the fact that the external field, $\omega_L^{z}$, adds to the internal spin-orbit field $\Omega_{\s SO}$ in the valley $K$ and $-\Omega_{\s SO}$ in the valley $K'$, see Fig. \ref{fig1}.

The system Eq. (\ref {general}) should be solved with initial conditions, ${\bm S}(0)={\hat x}$, which reflects the fact that at the moment of injection the spin is directed along the $x$-axis. In our earlier paper Ref. [\onlinecite{25}] we have demonstrated that the analytical solution
of the system depends on the dimensionless ratio
\begin{equation}
\Gamma=\frac{\gamma_v}{\Omega_{\s SO}}.
\end{equation}
For a slow intervalley scattering, $\Gamma <1$, the solution for the $x$ projection of the
net spin,  $S_x(t)=S_x^K(t)+S_x^{K'}(t)$, reads
\begin{widetext}
 \begin{align}
\label{spindensity}
S_x(t)&=\frac{1}{2}\Bigg\{\frac{\Gamma}{\sqrt{1-\Gamma^2}}\Bigg[\sin \Big(\omega_L^z + \tilde{\Omega}_{\s SO}\Big) t
- \sin\Big( \omega_L^z -\tilde{\Omega}_{\s SO}\Big)t \Bigg]
 \nonumber \\
+& \Bigg[\cos \Big(\omega_L^z + \tilde{\Omega}_{\s SO}\Big)t
+ \cos \Big(\omega_L^z -\tilde{\Omega}_{\s SO}\Big)t\Bigg]\Bigg\}\exp\Big[-\Gamma \Omega_{\s SO}t\Big],
 \end{align}
 \end{widetext}
where we have introduced modified spin-orbit coupling
\begin{equation}
\tilde{\Omega}_{\s SO}=\sqrt{1-\Gamma^2}\Omega_{\s SO}.
\end{equation}

From Eq. (\ref{spindensity}) the calculation of nonlocal resistance is straightforward and the result can be obtained in a closed form. This is apparent, because, in the absence of spin-orbit coupling, the integral of the product $\exp(i\omega_L^z t)P_d(t)$ appears in the expression for the conventional Hanle shape and can be evaluated analytically. In our case, $S_x(t)$ is the combination of two oscillating functions. Final result for  $R(\omega_L^{z})$ reads

\begin{widetext}
\begin{align}
\label{smallgamma}
&R(\omega_L^{z})= R_0 \Bigg\{\Big(\frac{\pi}{|y_+|}\Big)^{1/2}\exp\Big[-2|y_+|^{1/2}\cos\phi_+\Big]\cos\Big(\phi_+ +2|y_+|^{1/2}\sin\phi_+\Big)\nonumber\\
+&\Big(\frac{\pi}{|y_-|}\Big)^{1/2}\exp\Big[-2|y_-|^{1/2}\cos\phi_-\Big]\cos\Big(\phi_- +
2|y_-|^{1/2}\sin\phi_-\Big)\nonumber\\
-&\frac{\Gamma}{\sqrt{1-\Gamma^2}}\Bigg\{
\Big(\frac{\pi}{|y_+|}\Big)^{1/2}\exp\Big[-2|y_+|^{1/2}\cos\phi_+\Big]\sin\Big(\phi_+ +
2|y_+|^{1/2}\sin\phi_+\Big)\nonumber\\
-&\Big(\frac{\pi}{|y_-|}\Big)^{1/2}\exp\Big[-2|y_-|^{1/2}\cos\phi_-\Big]\sin\Big(\phi_- +
2|y_-|^{1/2}\sin\phi_-\Big)\Bigg\} \Bigg\}.
\end{align}
\end{widetext}
$R_0$ is the prefactor which depends on spin injection/detection polarization, channel dimensions, and material resistivity [\onlinecite{R_0}]. Parameters $y_{\pm}$ and $\phi_{\pm}$ entering into Eq. (\ref{smallgamma})
are defined as:
\begin{align}
\label{yandphismallgamma}
|y_{\pm}|&={\tilde d}^2\Big[1+\Big(\omega_L^z\pm\tilde{\Omega}_{SO}\Big)^2(\tau_s^\ast)^2\Big]^{1/2},\nonumber \\
\phi_{\pm}&=\frac{1}{2}\arctan\Big[\Big(\omega_L^z\pm\tilde{\Omega}_{SO}\Big)\tau_s^\ast\Big],
\end{align}
where ${\tilde d}$ is the dimensionless distance
\begin{equation}
\tilde d=\frac{d}{\lambda_s}=\frac{d}{(D\tau^*_s)^{1/2}},
\end{equation}
and $\tau_s^*$ is the inverse spin relaxation rate.
In the absence of other relaxation mechanisms, this rate is determined by the intervalley scattering rate. In the presence of additional mechanisms, this rate is the sum of the partial rates

\begin{equation}
\label{smallgammatau}
\frac{1}{\tau_s^{*}}=\Gamma{\Omega}_{\s SO}+\frac{1}{\tau_s}.
\end{equation}

\begin{figure}
\includegraphics[width=91mm]{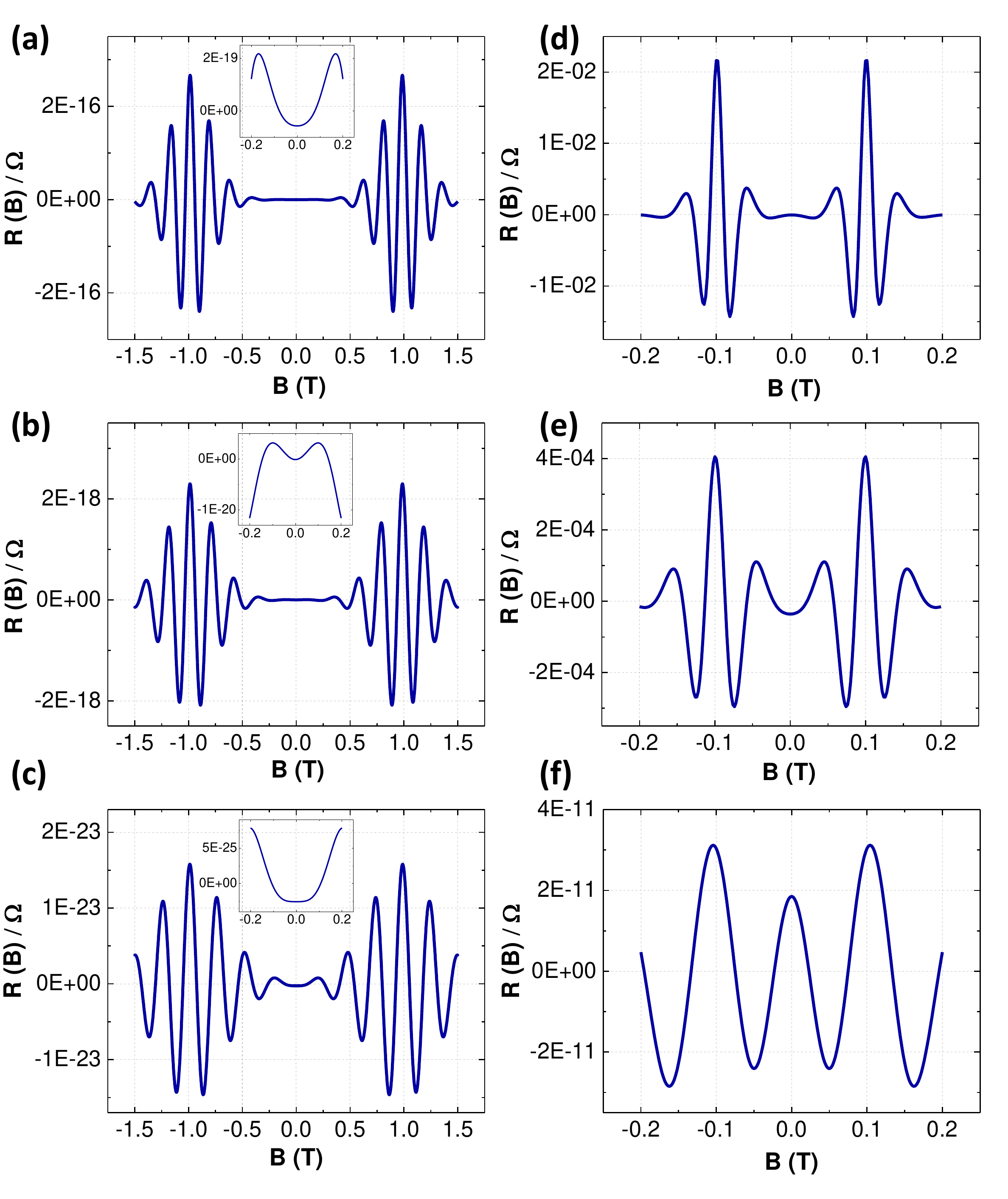}
\caption{Theoretical Hanle curve calculated from Eq. (\ref{smallgamma}). In all four panels the value of parameter $\Gamma$ is $0.2$, the distance $d$ is chosen to be $d=1$ $\mu$m, and the diffusion coefficient is equal to $D=2.6$ cm$^2/$s; (a) $B_{\s SO}$ = 1~T, $\tau_s^{*}$ = 10~ps, $\tilde d$ =~19.7; (b) $B_{\s SO}$ = 1~T, $\tau_s^{*}$ = 8~ps, $\tilde d$ =~22.0; (c) $B_{\s SO}$ = 1~T, $\tau_s^{*}$ = 5~ps, $\tilde d$ =~27.8; (d) $B_{\s SO}$ = 0.1~T, $\tau_s^{*}$ = 200~ps, $\tilde d$ =~4.4; (e) $B_{\s SO}$ = 0.1~T, $\tau_s^{*}$ = 100~ps, $\tilde d$ =~6.2; (f) $B_{\s SO}$ = 0.1~T, $\tau_s^{*}$ = 20~ps, $\tilde d$ =~13.9. Inset figures of (a-c) exhibit the zoom-in curve in smaller field range. }
\label{fig4}
\end{figure}

\subsection{The case of slow intervalley scattering}

In Fig. \ref{fig4} the Hanle curves plotted using Eq. (\ref {smallgamma})
for two different values of  spin-orbit field and several sets of $\tau_s^*$ are shown. For these plots, the value of
parameter $\Gamma$ was chosen to be $\Gamma=0.2$.
As can be seen, the nonlocal resistance decays away from $B=\pm B_{\s SO}$ in an oscillatory fashion. Naturally, for shorter $\tau_s^*$, the decay of oscillations is faster.
For ``strong" $B_{\s SO}=1$ T (as in a monolayer)
the most pronounced oscillations take place away from the origin.
However, with regard to experiment, we are interested in the behavior of nonlocal resistance
only within the domain $|B|<0.2$ T. For this reason, the central regions of the plots are enlarged. We see that the evolution of the Hanle shape
near $B=0$ is quite lively, so that  the shape changes significantly even when $\tau_s^*$
changes slightly from $\tau_s^* =5$ ps and $\tau_s^* =8$ ps. Still, the distance between
the two maxima exceeds $0.2$ T for all $\tau_s^*$ near $B=0$. For a smaller value of the spin-orbit field, $B_{\s SO}=0.1$ T
the behavior of nonlocal resistance near $B=0$ evolves with increasing $\tau_s^*$ as follows.
There are pronounced oscillations at $\tau_s^* =20$ ps, less pronounced at $\tau_s^* =100$ ps,
and almost no oscillations for $\tau_s^* =200$ ps. This behavior is the consequence of the fact
that the bigger is $\tau_s^*$, the more ``bound" are the oscillations to the points $B =\pm 0.1$ T.
Still, in all three curves the distance between the left and right extrema is close to $0.1$ T near $B=0$.


\subsection{The case of fast intervalley scattering}

 For  fast intervalley scattering we have $\Gamma > 1$ and the expression Eq. (\ref{smallgamma}) for spin dynamics
 does not apply anymore. Physically, in the domain  of fast intervalley scattering spin-orbit field effectively
 averages out as a result of fast switching of a carrier between the valleys.
 The modes of spin dynamics in the domain $\Gamma >1$ are classified into valley-symmetric (we denote it with $+$)
 and valley-asymmetric ($-$). As a result of averaging out of $\pm B_{\s SO}$ the $-$ mode has a long lifetime,
 $\sim \Omega_{\s SO}^2/\gamma_v$, while the lifetime of the symmetric mode is $\sim \gamma_v$.
 The actual form of $S_x(t)$ in the domain $\Gamma>1$ is still the sum of the  products of oscillating and
 exponentially decaying functions, as we have demonstrated in Ref. [\onlinecite{25}].
 This allows to calculate the Hanle curve explicitly for finite separation, $d$, between the contacts.
 The result, representing the sum of contributions from $+$ and $-$ modes, reads

\begin{widetext}
\begin{align}
\label{largegamma}
&R(\omega_L^{z})= R_0 \Bigg\{ \Big(1-\frac{\Gamma}{\sqrt{1-\Gamma^2}}\Big)
\Big(\frac{\pi}{|Y_+|}\Big)^{1/2}\exp\Big[-2|Y_+|^{1/2}\cos\Phi_+\Big]
\cos\Big(\Phi_+ +2|Y_+|^{1/2}\sin\Phi_+\Big)\nonumber\\
+&\Big(1+\frac{\Gamma}{\sqrt{1-\Gamma^2}}\Big)\Big(\frac{\pi}{|Y_-|}\Big)^{1/2}
\exp\Big[-2|Y_-|^{1/2}\cos\Phi_-\Big]\cos\Big(\Phi_- +
2|Y_-|^{1/2}\sin\Phi_-\Big)\Bigg\}.
\end{align}
\end{widetext}
The notations in Eq. (\ref{largegamma}) are the following
\begin{align}
\label{yandphilargegamma}
|Y_{\pm}|&={\tilde d_{\pm}}^2\Big[1+(\omega_L^z)^2(\tau_{s\pm}^\ast)^2\Big]^{1/2},\nonumber \\
\Phi_{\pm}&=\frac{1}{2}\arctan\Big[\omega_L^z\tau_{s\pm}^\ast\Big].
\end{align}
The relaxation times in Eq. (\ref{yandphilargegamma}) are defined as
\begin{equation}
\label{largegammatau}
\frac{1}{\tau_{s\pm}^*}=(\Gamma\pm\sqrt{\Gamma^2-1})\Omega_{\s SO}+\frac{1}{\tau_s}.
\end{equation}
Two values of $\tau_s^*$ result in two spin diffusion lengths, $\lambda_{s\pm}$, so that ${\tilde d}_{+}$ in
Eq. (\ref{yandphilargegamma}) is equal to $d/\lambda_{s+}$ and  ${\tilde d}_{-}$ is equal to  $d/\lambda_{s-}$.



\begin{figure}
\includegraphics[width=92mm]{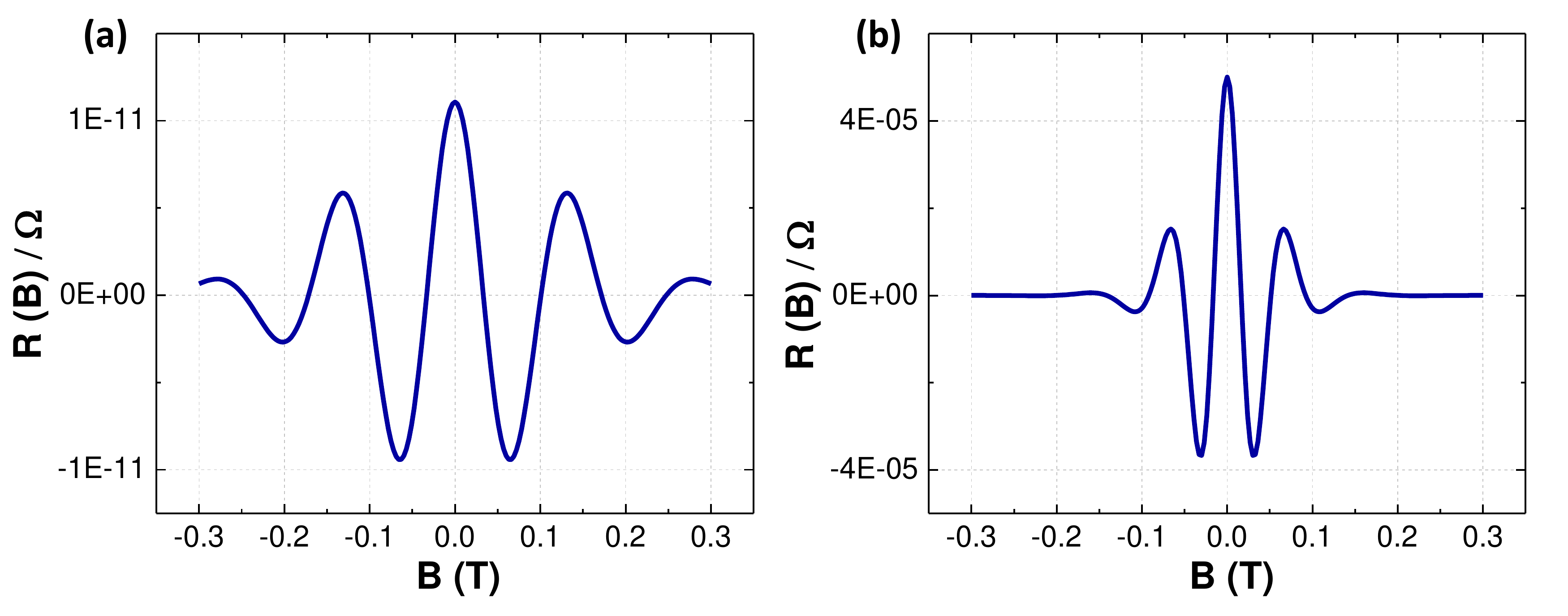}
\caption{Theoretical Hanle curve calculated from Eq. (\ref{largegamma}), $R(B)$ vs external field $B$, under conditions of $\Gamma = 2$ for different $B_{\s SO}$ and ${\tau_s^{*}}_{\pm}$. (a) $B_{\s SO}$ = 0.1~T, $\tau_{s+}^\ast$ = 3~ps, $\tau_{s-}^\ast$ = 68~ps, $\tilde d_{+}$ =~17.1, $\tilde d_{-}$ =~7.5; (b) $B_{\s SO}$ = 1~T, $\tau_{s+}^\ast$ = 1.5~ps, $\tau_{s-}^\ast$ = 18~ps, $\tilde d_{+}$ =~50.8, $\tilde d_{-}$ =~14.9. }
\label{fig5}
\end{figure}

Fig. \ref{fig5} shows the resulting Hanle curves calculated from Eq. (\ref{largegamma}) for different values of $B_{\s SO}$. A distinctive feature of these curves compared to Fig. \ref{fig4} is that $R(B)$ falls off from $B=0$ with oscillations, so  that the maximum at $B=0$ is the highest.

%

\section{EXPERIMENTAL HANLE DATA}
\begin{figure}
\includegraphics[width=70mm]{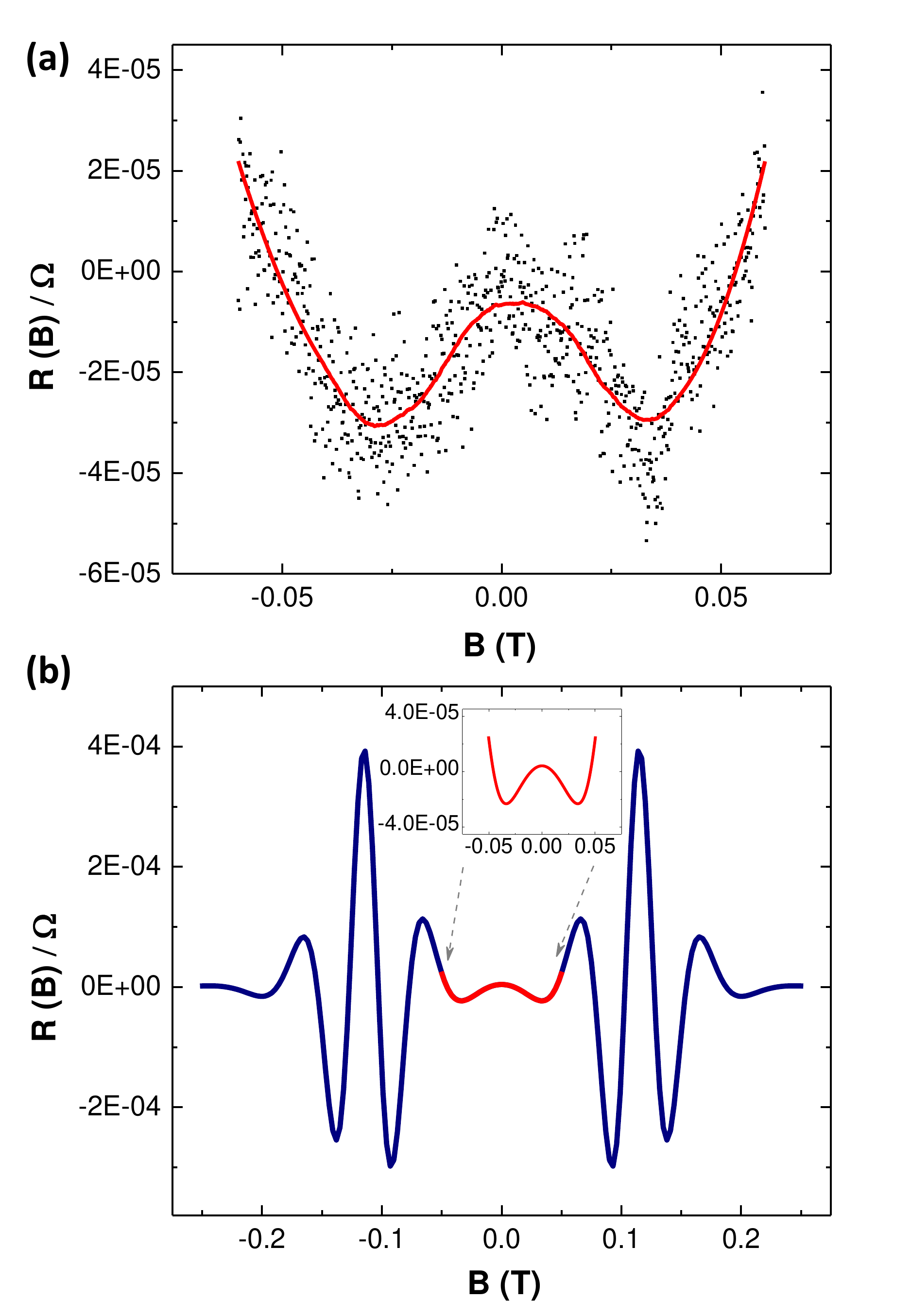}
\caption{(a) Experiment results, $R(B)$, in normal field (maximum 0.06~T) performed at 30~K. The red curve is the guide to eye line. (b) Calculated $R (B)$ curve using Eg. (\ref{smallgamma}) which the zoom-in portion of the curve matches the experiment result in (a) as shown in the red scatter part of the complete Hanle curve. The fitting parameters are as follows: $\mu=900~cm^2/Vs$, $\Gamma = 0.35 $, $B_{\s SO} = 0.12~T$, $\tau_s^{*} = 110~ps$, $\tilde d = 6.25$. The observations indicate possibility of an oscillatory signal present in the weaker field of normal direction using prediction of small intervalley scattering rate. }
\label{fig6}
\end{figure}

Fig. \ref{fig6} shows the experimental Hanle data recorded at $30~K$. Measurements of nonlocal resistance were performed using the experimental geometry shown in Fig. \ref{fig2}(b). Before measurements, injector and detector contacts were magnetized parallel to each other by applying an in-plane magnetic field which was parallel to the length of the electrode. Once the electrodes were magnetized, in-plane field was removed and a perpendicular magnetic field was applied to record Hanle data. To avoid the misalignment of the injector and detector magnetization, the out-of plane magnetic field was restricted between $\pm60$ mT. For Hanle measurements, current was applied between electrode 1 and 2, and the voltage was measured between 3 and 4 while the transverse magnetic field was scanned from $+60$ mT to $-60$ mT.

Some salient features of the observed $R_{\s NL}$ vs. $B$ curve are: (a) value of $R_{\s NL}$ at $B= 0$ T is almost zero; (b) on increasing the magnetic field, the sign of $R_{\s NL}$ becomes negative and its magnitude starts increasing on either side of $B = 0$ T; (c) at around $B = 0.035$ T, the magnitude of $R_{\s NL}$ starts decreasing and finally at $B = 0.05$ T, again becomes zero; (d) after $0.05$ T, $R_{\s NL}$ become positive and its magnitude increases with increase in field. The above features are obviously very different from the typical Hanle curves shown by normal materials where a maxima is observed at $B = 0$ T in $R_{\s NL}$ vs. $B$ curve. In order to understand the mechanism responsible for the observed behavior, we compare the experimental data with our theoretical prediction discussed in section III.

First of all, a quick comparison of the curve shown in Fig. \ref{fig6}(a) with theoretical curves shown in Fig. \ref{fig4} and Fig. \ref{fig5}, suggested that the sample under investigation is in the regime $\Gamma < 1$. Specifically if the sample was in the regime $\Gamma > 1$, it should have exhibited a maxima at $B = 0$ T in $R_{\s NL}$ vs. $B$ curve. Once we found out the regime to which our sample belongs, we fitted the experimental data to the corresponding expression, i.e. equation (\ref{smallgamma}). In equation (\ref{smallgamma}), there are four independent unknown parameters namely, $B_{\s SO}$, $\tau_s^{*}$, $\Gamma$, and $\mu$, all of which were varied during the fitting procedure. Figure of merit of the fit was determined by calculating the quantity $\chi^{2} = \frac{1}{N} \sum_{n=1}^{N}[\frac{Y_{data} - Y_{fit}}{Y_{error}}]^2$.

The experimental data was found to fit very well in equation (7) with a  value of $\chi^{2}= 1.16$. Fitted curve is shown by solid red curve in Fig. 6(a). The best fitting parameters were found to be $B_{\s SO} = 0.12\pm0.01$ T, $\tau_s^{*} = 110\pm10~ps$, $\Gamma = 0.35\pm0.05$, and $\mu = 900\pm50~{cm}^2/Vs$.

It is important to note that even though, because of experimental constraints, range over which we could scan the out-of-plane field was limited to $\pm 60$ mT, we could estimate $B_{\s SO}$ which was much higher than that field. Now using the experimental determined values of $B_{\s SO}$, $\tau_s^{*}$, $\Gamma$, and $\mu$,  in equation (\ref{smallgamma}), we calculated the $R(B)$ curve over a magnetic field range of $\pm0.25$ T and obtained a curve shown in Fig. \ref{fig6}(b).

The curve exhibits two peak structure with maxima located at $\pm0.12$ T corresponding to two valleys. Furthermore, both of the peaks are accompanied with oscillatory signal on both sides of the main peaks which is understood to arise because of the integer number of full precession accomplished by the spin of the injection electrode before it reaches the detector electrode.

It is interesting to note that though the main Hanle peaks belonging to two valleys are well separated… corresponding oscillatory peaks overlap near the origin and give rise to the shape in $R_{\s NL}$ vs. $B$ plot as observed in our experiment. See the part of curve shown in red line in Fig. \ref{fig6}(b).

To check the consistency of the fit, from the values of $\Gamma$ and $B_{\s SO}$ obtained above by the fitting of experimental data, we calculate the intervalley scattering rate $\gamma_v=\Gamma B_{\s SO}=7.4\times{10}^{9}~s^{-1}$. If the mobility is limited entirely by the intervalley
scattering, it is related to $\gamma_v$ and the carrier density $n$ as [\onlinecite{25}]
\begin{equation}
\label{gammaratio}
\mu=\left(\frac{\Delta}{\lambda}\right)\frac{e}{2\pi\hbar \Gamma n}=\left(\frac{\Delta}{\lambda}\right)\frac{e B_{\s SO}}{2\pi\hbar \gamma_v n}.
\end{equation}

Using the value of mobility obtained from the fit, and the  value of density $n = 10^{13}~cm^{-2}$ from  Ref. [3] we got the value $\Gamma =0.20$ in reasonable agreement with $\Gamma =0.35$ inferred from the fit.

\section{CONCLUDING REMARKS}

\noindent{(i)} In summary, using the all-electrical technique of injecting and detecting spin polarized carriers, we have observed the signature of Hanle precession in trilayer MoS$_2$ films.

\noindent{(ii)} Our theoretical calculations showed that because of the valley-specific spin-orbit field present in the odd-layered MoS$_2$ films, two distinct Hanle peaks centered at $B = \pm B_{\s SO}$ are expected.

\noindent{(iii)} In the case of trilayer MoS$_2$, the strength of $SO$ field is much smaller than that for monolayer films. As a result, under certain experimental conditions, secondary oscillatory signals belonging to the two valleys can overlap and give rise to a detectable signal near the zero external magnetic field.

\noindent{(iv)} By comparing the experimental data with the theoretically predicted results, we found that the trilayer MoS$_2$ films prepared by PLD undergo a slow intervalley scattering which is very important from the point of view of realizing practical valleytronic devices. A spin life-time of around $110~ps$ was estimated at $30$ K.

\noindent{(v)} Here, it is also very instructive to compare the results of our all-electrical study with the study reported in [\onlinecite{Crooker}] where the optical techniques were employed for spin injection and detection in monolayer MoS$_2$. To achieve the optical response the authors had to apply the normal magnetic field as high as $65$ T. In the present study, we observed the sensitivity of the spin transport in trilayer MoS$_2$ to much smaller fields. One reason for this observed difference is the fact that in trilayer MoS$_2$, $SO$ field is much smaller than that in monolayer MoS$_2$. The other reason, more important from the point of view of spin-transport, is that in our present transport measurement geometry, the injector and detector were separated by distance much longer than the spin diffusion length. As a result, the nonlocal voltage is created not by typical electrons, which loose their spin memory after time $\tau_s^{*}$, but by electrons that escape relaxation for a long time before reaching the detector. Precession of these electrons is substantial in much weaker fields. As a result, the Hanle signal is weak, but still distinguishable.

\noindent{(vi)} Availability of large-area trilayer MoS$_2$ films, in which valley specific spin transport can be investigated by electrical means, is likely to expedite further research in this area.

\section{ACKNOWLEDGEMENTS}
This work was supported by NSF through grant No. 1407650 and 1121252.


\appendix
\section{Calculation of Schottky barrier height using thermionic equation}
 In the case of 2D materials, the thermionic emission equation is [\onlinecite{27}]:
\begin{equation}
I_{12}=A_{2D} S T^{3/2} \exp \Big[-\frac{q}{k_B T}\Big(\Phi_B - \frac{V_{12}}{n}\Big)\Big],
\end{equation}
where $A_{2D}$ is the $2$D equivalent Richardson constant, $S$ is the contact area between MoS$_2$ film and FM probe, $q$ is electron charge, $n$ is the ideality factor, $k_B$ is the Boltzmann constant.  The slope of the Arrhenius plot, $In(I_{12}/T^{3/2})$ vs $1000/$T, is given by the expression:
\begin{equation}
S=- \frac {q}{1000 k_B}\Big(\Phi_B - \frac{V_{12}}{n}\Big).
\end{equation}


\begin{thebibliography}{99}
\bibitem{R_0}
We define $R_0 = P^{2}\rho d/(8\sqrt{\pi}A)$. For our trilayer MoS$_2$ system, we calculated $R_0$ is about 350 $\Omega \cdot m/s$ assuming $P$ is 0.7.
\end{thebibliography}
\end{document}